\documentclass{article}
\usepackage{spconf,amsmath,graphicx}

\usepackage{amsfonts}
\usepackage{multirow}
\usepackage{hyperref}
\usepackage{subfigure}
\usepackage{booktabs,setspace}
\usepackage{threeparttable}
\ninept
\usepackage{setspace}

\usepackage{amssymb,CJK, indentfirst,balance,appendix}

\title{SUMMARY ON THE ICASSP 2022 MULTI-CHANNEL MULTI-PARTY MEETING TRANSCRIPTION GRAND CHALLENGE}

\name{\begin{tabular}{c}Fan Yu$^{1,3}$, Shiliang Zhang$^{1}$, Pengcheng Guo$^3$, Yihui Fu$^3$, Zhihao Du$^1$, Siqi Zheng$^1$, Weilong Huang$^1$, Lei Xie$^{3*}$\thanks{* Lei Xie is the corresponding author.}, \\
Zheng-Hua Tan$^5$, DeLiang Wang$^6$, Yanmin Qian$^7$,  Kong Aik Lee$^8$, Zhijie Yan$^1$, Bin Ma$^2$, Xin Xu$^4$, Hui Bu$^4$\end{tabular}}

\address{$^1$Speech Lab, Alibaba Group, China 
        $^2$Speech Lab, Alibaba Group, Singapore \\
        $^3$AISHELL Foundation 
        $^4$Beijing Shell Shell Technology Co., Ltd., Beijing, China \\
        $^5$Department of Electronic Systems, Aalborg University, Aalborg, Denmark \\
        $^6$Department of Computer Science and Engineering, \\
        Center for Cognitive and Brain Sciences, The Ohio State University, USA \\
        $^7$SpeechLab, Department of Computer Science and Engineering, Shanghai Jiao Tong University, China \\
        $^8$Aural and Language Intelligence Department, Institute for Infocomm Research, A*STAR, Singapore
        }

\begin{document}

%
\maketitle

\begin{abstract}
\vspace{-0.1cm}

The ICASSP 2022 Multi-channel Multi-party Meeting Transcription Grand Challenge (M2MeT) focuses on one of the most valuable and the most challenging scenarios of speech technologies. The M2MeT challenge has particularly set up two tracks, speaker diarization (track 1) and multi-speaker automatic speech recognition (ASR) (track 2). Along with the challenge, we released 120 hours of real-recorded Mandarin meeting speech data with manual annotation, including far-field data collected by 8-channel microphone array as well as near-field data collected by each participants' headset microphone. We briefly describe the released dataset, track setups, baselines and summarize the challenge results and major techniques used in the submissions. 


\end{abstract}

\begin{keywords}
M2MeT, Alimeeting, Meeting Transcription, Speaker Diarization, Multi-speaker ASR
\end{keywords}
\vspace{-0.4cm}
\section{Introduction}
\vspace{-0.3cm}
\textit{Who spoke what at when} is the major aim of rich transcription of real-world multi-speaker meetings. Despite years of research~\cite{fiscus2005rich,fiscus2006rich,fiscus2007rich}, meeting rich transcription is still considered as one of the most challenging tasks in speech processing due to free speaking styles and complex acoustic conditions, such as overlapping speech, unknown number of speakers, far-field attenuated speech signals in large conference rooms, noise, reverberation, etc. As a result, tackling the problem requires a well-designed speech system with multiple related speech processing components, including but not limited to front-end signal processing, speaker identification, speaker diarization and automatic speech recognition (ASR).



The recent advances of deep learning has boosted a new wave of related research on meeting transcription, including speaker darization~\cite{park2022review,fujita2019end2,horiguchi2020end}, speech separation~\cite{yu2017permutation,hershey2016deep,chen2017deep} and multi-speaker ASR~\cite{yu2017recognizing,chen2017progressive,kanda2020serialized}. The ICASSP 2022 Multi-Channel Multi-Party Meeting Transcription Challenge (M2MeT) \footnote{Challenge website: https://www.alibabacloud.com/m2met-alimeeting} was designed with the aim to provide a common evaluation platform and a sizable dataset for Mandarin meeting transcription~\cite{yu2021m2met}. Along with the challenge, we made available the \textit{AliMeeting} dataset to the participants, which contains 120 hours real meeting data recorded by 8-channel directional microphone array and headset microphone. Two tracks are particularly designed. Track 1 is speaker diarization, in which participants are tasked with addressing the “who spoke when” question by logging speaker-specific speech events on multi-speaker audio data. Track 2 focuses on transcribing multi-speaker speech that may contain overlapped segments from multiple speakers.


This paper summarizes the challenge outcomes. Specifically, we give a brief literature overview on speaker diarization and multi-speaker ASR in Section 2. Section 3 reviews the released dataset and the associated baselines. Sections 4 and 5 discuss the outcome of the challenge with major techniques and tricks used in submitted systems. Section 6 concludes the paper.

\vspace{-0.45cm}
\section{Related Works}
\vspace{-0.3cm}
Speaker diarization and multi-speaker ASR in meeting scenarios have attracted increasing attention. 
For speaker diarization, conventional clustering-based approaches usually contain a speaker embedding extraction step and a clustering step where the input audio stream is first converted into speaker-specific representation~\cite{snyder2018x}, followed by a clustering process, such as Variational Bayesian HMM clustering (VBx)~\cite{landini2022bayesian}, which aggregates the regions of each speaker into separated clusters. The clustering based approach is ineffective to recognize the overlapped speech without additional modules, because it assumes that each speech frame corresponds to only one of the speakers. Therefore, resegmentation was used by~\cite{sell2015diarization} to handle the overlap segments. However, overlapping speech detection (OSD)~\cite{bredin2020pyannote} is also a challenging task itself in most situations. Compared with majority studies that work on two-talker telephony conversations, there is a recent trend to handle more challenging speaker dairzation scenarios in complicated talking and acoustic environments~\cite{watanabe2020chime,barker2018fifth,ryant2019second,ryant2020third}. For the multi-speaker meetings recorded with a microphone array from distance, the scenario considered in this challenge, speaker darization becomes more challenging as speaker overlaps happen more frequently and sometimes several speakers speak at the same time in a conference discussion.

The advances of deep learning have shed light on the problem. As a typical solution, recent end-to-end neural diarization (EEND)~\cite{fujita2019end2} and its variants~\cite{horiguchi2020end} have replaced the individual sub-modules in traditional speaker diarization systems mentioned above with one neural network that directly provides the overlap-aware diarization results. More promisingly, thanks to the advances of speaker embedding extraction~\cite{kanagasundaram2011vector,snyder2018x}, target speaker vocie actvity detecion (TS-VAD)~\cite{medennikov2020target,he2021target} was proposed to judge target speaker’s activeness for each speech frame, which can estimate multiple speakers at the same time, leading to a promising solution to handle overlapped speech.


With the development of deep learning, end-to-end neural approaches have rapidly gained prominence in the speech recognition community~\cite{li2021recent}.
However, ASR in complicated scenarios such as meetings is still not a solved problem with challenges including complex acoustic conditions, unknown number of speakers and overlapping speech.
In other words, the challenges mentioned above in speaker diarization also exist in multi-speaker ASR.
Besides multi-condition training and data augmentation, speech enhancement~\cite{hu2020dccrn} and separation~\cite{luo2019conv,liu2020causal} are considered as remedy to the complex acoustic conditions and multiple speakers. Speech enhancement that explicitly addresses the background noise has been widely studied~\cite{watanabe2020chime,heymann2017beamnet,koizumi2021snri}, where multi-channel signals can be adopted if microphone array is deployed for audio recording. Speech separation and joint-training with ASR under permutation invariant training (PIT) scheme were studied to achieve a high-performance ASR system for overlapped speech~\cite{chen2020continuous}. Designing an end-to-end system that directly outputs multi-speaker transcriptions seems a straightforward solution to multi-talker ASR, such as the multi-channel input multi-speaker output (MIMO) approach~\cite{chang2019mimo,zhang2021end} and the end-to-end unmixing, fixed-beamformer and extraction (E2E-UFE) system~\cite{wu2020end}. Conditional chain~\cite{shi2020sequence} was also proposed to solve the PIT problem of the number of speakers is unknown.
However, the above approaches rely on complicated joint training of front and back-end models or re-designing a complicated neural architecture. The SOT method~\cite{kanda2020serialized} does not change the original ASR network structure designed for single speaker. Instead, it only introduces a special symbol to represent the speaker change.
Moreover, speech recognition and diarization for unsegmented multi-talker recordings with speaker overlaps was discussed in the recent JSALT workshop to further promote reproducible research in this field~\cite{raj2021integration}.


Different from relevant datasets that have been released before~\cite{watanabe2020chime,janin2003icsi,mccowan2005ami}, AliMeeting released in this challenge and Aishell-4~\cite{fu2021aishell} are currently the only publicly available meeting datasets in Mandarin. Specifically, AliMeeting has more speakers and meeting venues, while particularly adding multi-speaker discussions with a high speaker overlap ratio.

\vspace{-0.4cm}
\section{DATASETS, TRACKS AND BASELINES}
\vspace{-0.2cm}
As described in our challenge evaluation plan~\cite{yu2021m2met}, AliMeeting contains 118.75 hours~\footnote{Hours are calculated in single channel of audio.} of speech data in total. The training set (Train) and evaluation set (Eval) are first released to participants for system development, with 104.75 and 4 hours of speech, respectively, with manual transcription and timestamp. During the challenge ranking period, the 10 hours test set (Test) is released for scoring. Specifically, the Train, Eval and Test sets contain 212, 8 and 20 meeting sessions respectively, and each session consists of a 15 to 30-minute discussion by 2-4 participants. To highlight speaker overlap, the sessions with 4 participants account for 59\%, 50\% and 57\% sessions in Train, Eval and Test, respectively. For Train and Eval sets, we provide the 8-channel audio recorded from the microphone array in far-field as well as the near-field audio from the participant's headset microphone, while the Test set only contains the 8-channel far-field audio.

The challenge consists of two tracks, namely speaker diarization (track 1) and multi-speaker ASR (track 2), measured and ranked on the Test set by Diarization Error Rate (DER) and Character Error Rate (CER) respectively. For both tracks, we also set up two sub-tracks. For the constrained data sub-track, system building for both tracks are restricted to AliMeeting~\cite{yu2021m2met}, Aishell-4~\cite{fu2021aishell} and CN-Celeb~\cite{fan2020cn}, while for the unconstrained data track, participants can use any data set publicly available.

We release baseline systems along with the Train and Eval data for quick start and reproducible research. For the 8-channel data of AliMeeting recorded by microphone array, we select the first channel to obtain \textit{Ali-far}, and adopt CDDMA beamformer~\cite{huang2020differential,zheng2021real} on 8-channel data to generate \textit{Ali-far-bf}. We use prefix \textit{Train-*}, \textit{Eval-*} and \textit{Test-*} to denote generated data associated with Train, Eval and Test sets. For example, \textit{Test-Ali-far-bf} means the beamformed data for the Test set.

We adopt the Kaldi-based diarization system from the CHiME-6 challenge as the baseline system for track 1. The diarizaiton module includes speaker embedding extractor and clustering. DER is scored with collar size of 0 and 0.25 second, but the challenge ranking is based on the 0.25 second collar size. The speaker diarization results for the baseline system are shown in Table~\ref{tab:speakers}.

\begin{table}[!htb]
\centering
\vspace{-0.6cm}
\caption{Speaker diarization results on Eval and Test in DER (\%).}
\vspace{0.15cm}
\begin{threeparttable}[t]
\begin{tabular}{lcc}
\toprule

Testing data        & Collar size = 0 & Collar size = 0.25 \\ \midrule
Eval-Ali-far    &   \textbf{24.52}     &    \textbf{15.24}    \\
Eval-Ali-far-bf &   24.67	    &     15.46        \\
Test-Ali-far    &    \textbf{24.95}      &   \textbf{15.60}       \\
Test-Ali-far-bf &    25.16      &   15.79               \\ 
\bottomrule
\end{tabular}
\vspace{-0.3cm}
\end{threeparttable}
\label{tab:speakers}

\end{table}

We use a Conformer-based~\cite{gulati2020conformer} ASR model as our single speaker baseline (ConfomerA), which is trained by \textit{Train-Ali-near} set using ESPnet~\cite{watanabe2018espnet}. We adopt Serialized Output Training (SOT)~\cite{kanda2020serialized} to recognize speech from multiple speakers containing overlapped speech, generating transcriptions of multiple speakers one after another. The baseline results of multi-speaker ASR are shown in Table~\ref{tab:asr_multi}. Note that here SOT and SOT\_bf are trained by \textit{Train-Ali-near} and \textit{Train-Ali-far} respectively. Compared with the single-speaker conformer model (ConfomerA), the two SOT multi-speaker models have obtained significant improvement on the Eval and Test sets, where SOT\_bf achieves superior performance. 

More details on the data arrangements, tracks and baseline results can be referred to the challenge evaluation plan paper~\cite{yu2021m2met}.

\begin{table}[!htb]
\centering
\vspace{-0.6cm}
\caption{Multi-speaker ASR results on Eval and Test in CER (\%).}
\vspace{0.15cm}
\begin{tabular}{lccc}
\toprule
Testing data               & ConfomerA       & SOT             & SOT\_bf     \\ \midrule
Eval-Ali-far           & 49.0    & 30.8   & 34.3          \\
Eval-Ali-far-bf        & 45.6    & 33.2            & \textbf{29.7}  \\
Test-Ali-far           & 50.4    & 32.4   & 35.9              \\
Test-Ali-far-bf        & 46.3   & 33.9            & \textbf{30.9}    \\
\bottomrule
\end{tabular}
\label{tab:asr_multi}
\vspace{-0.6cm}
\end{table}

\begin{table*}[htbp] 
	\caption{Top 8 ranking teams in terms of DER in track 1 and their major techniques.}
	\vspace{0.1cm}
	\centering 
	\includegraphics[scale=0.46]{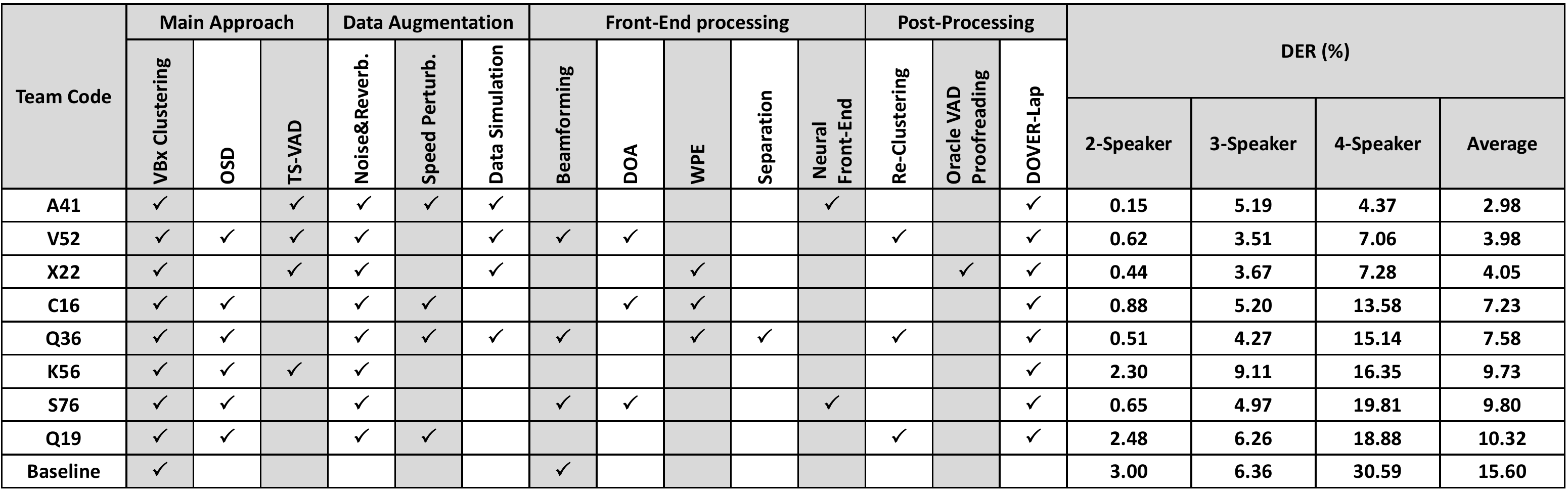} 
	\label{track1}
	\vspace{-0.6cm}
\end{table*}

\vspace{-0.2cm}
\section{Summary on Track 1 - Speaker Diarization}
\vspace{-0.3cm}
Finally 14 teams submitted their results to track 1 and the DER for the top 8 teams is summarized in Table~\ref{track1}. Observing the performance by the number of speakers, we can see that in general, the DER increases with the number
of speakers in meeting sessions. For most teams, there are clear performance gaps between 2- and 3-speaker sessions and between 3- and 4-speaker sessions.
The winner goes to team A41~\cite{wang2022crosschannel} which achieves the lowest DER of 2.98\%, surpassed the official baseline (15.60\%) with a large margin. Interestingly, their system works equally well on both 3- and 4-speaker sessions. 
There are two key techniques ensuring their superior performance: using TS-VAD~\cite{medennikov2020target} to find speaker overlap and employing cross-channel self-attention~\cite{chang2021multi} to further improve performance. Table~\ref{track1} also summarizes the major techniques used by the top 8 teams, namely effective main approach, data augmentation strategy, front-end processing as well as post-processing. We will highlight these key techniques in the following.



\vspace{-0.5cm}
\subsection{Main Approach}
\vspace{-0.2cm}
With the assumption that each speech frame corresponds to only one speaker, a clustering-based speaker diarization system is incapable of handling overlapped speech without additional modules. Since AliMeeting has a high ratio of speaker overlap, it is beneficial to adopt effective methods to reduce the error brought by the overlapped speech. 
The top three teams all employ TS-VAD to find the overlap between speakers while end-to-end approach, e.g., EEND, is not considered. We believe that this is because the single-speaker speech segments in meeting recordings can be effectively used (through clustering) to obtain speaker embedding as the initial input for the TS-VAD model that has been proven consistently effective for handling overlapped speech in the literature.
Instead of using the original TS-VAD that takes i-vector as target-speaker embedding, the winner team A41~\cite{wang2022crosschannel} uses the deep speaker embedding extracted by ResNet~\cite{he2016deep} to detect the target-speaker. Moreover, with the premise that different acoustic features are complementary, the second-place team V52~\cite{zheng2022cuhktencent} proposes a multi-level feature fusion mechanism for TS-VAD, and the fusion between spatial-related and speaker-related features leads to 2\% absolute DER reduction on Eval set. Some teams adopt approaches to improve the clustering-based algorithm itself. For example, it is effective to use overlap speech detection (OSD) to divide oracle VAD segments into single speaker segments and overlapped speech segments. Moreover, estimating the direction of arrival (DOA) to distinguish different speakers by the corresponding spatial information is proven to be beneficial. Team Q36~\cite{tian2022royalflush} demonstrates that re-assigning speaker labels to the overlapping segments by a speech seperation method can lead to 14.32\% relative DER reduction on Eval set (7.47\% to 6.40\%).
\vspace{-0.5cm}
\subsection{Data Augmentation}
\vspace{-0.2cm}
Since the size of the released training data is relatively small, data augmentation is adopted by most teams. For example, noise augmentation and reverberation simulation are generally used, which improves the robustness of the model modestly. Simulated room impulse response (RIR) is used to convolve with the original speech to generate data with reverberation. To further augment the training samples, Team A41, C16 and Q36 adopt the amplification and tempo (change audio playback speed but do not change its pitch) to audio signals. Moreover, as speaker overlap is salient in the data, several teams create an extra simulated dataset based on Alimeeting and CN-celeb. In detail, utterances from different speakers are randomly selected from these data, and then combined with an overlap ratio from 0 to 40\%. It is also worth noticing that the winner team A41 simulates data in an online manner in order to obtain more diverse data and stronger model robustness.

\vspace{-0.45cm}
\subsection{Front-End Processing}
\vspace{-0.2cm}
Front-end processing approaches, such as dereverberation, beamforming and speech enhancement, have proven to be effective for downstream tasks dealing with far-field speech. In the challenge, team X22~\cite{he2022ustcximalaya}, C16 and Q36 adopt the weighted prediction error (WPE) based on long-term linear prediction for dereverberation, leading to an absolute 0.7\% DER reduction on the Eval set. Moreover, the relevant experiments from team X22 show that the offline dereverberation mode is more effective than the online mode. Interestingly, team Q36 found that using multi-channel WPE is harmful to OSD while it is beneficial for speaker clustering and speech separation. Effective adoption of spatial information, including beamforming~\cite{anguera2007acoustic}, is also mainly considered by the participants. In particular, team S76 proposes a novel architecture named discriminative multi-stream neural network (DMSNet) for overlapped speech detection. Instead of adopting beamforming, the winner team A41 employs cross-channel self-attention to integrate multi-channel signals, where the non-linear spatial correlations between different channels are learned and fused.

\vspace{-0.4cm}
\subsection{Post-Processing}
\vspace{-0.2cm}
Since the challenge does not restrict on the computation workload and system fusion, most teams employ the DOVER-Lap~\cite{raj2021dover} to fuse multiple effective models. The improvement from DOVER-Lap fusion depends on the number and type of models, and the relative DER reduction ranges from 2\% to 15\%. Note that although conventional VBx clustering is not as good as TS-VAD, but it brings extra gain after model fusion. Re-clustering is also an effective method for conventional clustering-based speaker diarization, which is applied to further refine the number of speakers by combining the very similar clusters according to their cosine distances. Since our challenge provides oracle VAD, Team X22 fuses the results with oracle VAD by deleting wrong speech segments and labeling the silent segments, leading to 22.2\% relative DER reduction on Eval set (13.04\% to 10.14\%).

\begin{table*}[htbp] 
	\caption{Top 5 ranking teams in terms of CER in track 2 and their major techniques.}
	\vspace{0.1cm}
	\centering 
	\includegraphics[scale=0.48]{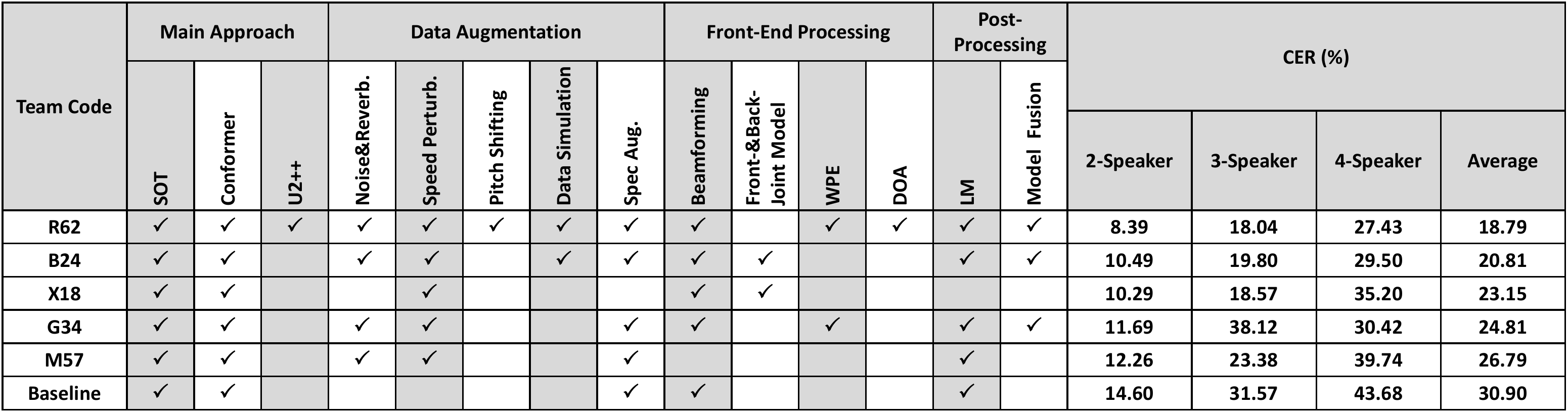} 
	\label{track2}
	\vspace{-0.6cm}
\end{table*}

\vspace{-0.4cm}
\section{Summary on Track 2 - Multi-speaker ASR}
\vspace{-0.3cm}

Twelve teams submitted their results to track 2 and the CER for the top 5 teams is summarized in Table~\ref{track2}. Similar to the observation in track 1, CER sharply increases with the number of speakers in the meeting sessions, mainly due to the high speaker overlap ratio in meetings with more speakers. The winner team R62~\cite{ye2022royalflush} obtained the lowest average CER of 18.79\% with over 12\% absolute CER reduction as compared with the baseline. The superior performance comes from a SOT-based multi-speaker ASR system with large-scale data simulation. Moreover, system fusion is also beneficial as reported by the winner team. In their approach, the standard conformer-based joint CTC/Attention Conformer~\cite{gulati2020conformer} and U2++~\cite{wu2021u2++} model with a bidirectional attention decoder are fused with clear performance gain. The main approaches, data augmentation strategies, front- and back-end processing are summarized in Table~\ref{track2}.

\vspace{-0.4cm}
\subsection{Main Approach}
\vspace{-0.2cm}
The 5 top teams all adopt the SOT approach~\cite{kanda2020serialized} similar to our multi-speaker baseline system, resulting in over 15\% CER reduction compared with single speaker ASR system on the Eval set. The SOT method has an excellent ability to model the dependencies among outputs for different speakers and no longer has a limitation on the maximum number of speakers. Undoubtedly, the Conformer architecture~\cite{gulati2020conformer}, which models both local and global context of speech, is 
used by all teams. It should be noted that besides Conformer, the winner team R62~\cite{ye2022royalflush} also uses the recent U2++~\cite{wu2021u2++} structure, where a bidirectional attention decoder is used to integrate information from both directions at inference. As a result, the fusion of the two models brings 8.7\% relative CER reduction on the Eval set.

\vspace{-0.4cm}
\subsection{Data Augmentation}
\vspace{-0.2cm}
Similar to track 1, various data augmentation tricks were widely adopted in track 2. Noise augmentation, reverberation simulation, speed perturbation and SpecAugmentation are the mainstream methods with stable performance improvement. According to the report provided by second-place team B24~\cite{shen2022volcspeech}, relative CER reduction of 13.5\% can be achieved by multi-channel multi-speaker data simulation as compared with the baseline trained using Train-Ali-far. Compared with speaker diarization, data simulation for multi-speaker ASR is more complex, which needs to consider various factors such as speaker turn and conversation duration. Thus fine-grained data simulation is essential to ensure consistent performance gain. For example, the simulation on speaker overlapping ratio should be reasonable, including the coverage of extreme cases like sudden (very brief) interruption from another speaker. It is also worth noticing that the winner team R62 makes substantial efforts in data augmentation and simulation. Finally, they expand the original training data to about 18,000 hours, which achieves 9.7\% absolute CER reduction compared with the baseline system.

\vspace{-0.4cm}
\subsection{Front-End Processing}
\vspace{-0.2cm}

Similar to track 1, the classical front-end processing techniques in far-field speech recognition, including beamforming, dereverberation and DOA, are also adopted in track 2 with performance gain. Specifically, beamforming is used by most teams and WPE-based dereverberation is considered by two teams, while DOA estimation of target speaker is only used by the winner team R62 among the top 5 teams.
Front-end and back-end joint modeling using neural networks is also considered by the second- and third-place team (B24 and X18). With the premise that optimizing front-end and back-end separately will lead to sub-optimal performance, joint modeling will make the whole system to be optimized under the final metric. Team B24 and X18 both take multi-channel signal as the input of a neural front-end and then cascade the front-end with the back-end Conformer ASR model. The whole neural architecture is then jointed trained. According to the report from B24, joint modeling leads to 13.3\% relative CER reduction (from 24.0\% to 20.8\%) on Eval set.

\vspace{-0.4cm}
\subsection{Post-Processing}
\vspace{-0.2cm}
As reported by several teams, the contribution from language modeling (LM), either $n$-gram or neural LM, is very weak. This is mainly because the building of LM is only restricted to the transcripts of the training data while using extra text data is prohibited according to the challenge rule. Most teams employ model fusion which brings absolute improvement ranging from 10\% to 15\% on the Eval set as compared with the baseline. For example, the winner team R62 has eventually fused 7 models by simple ROVER, including 3 Conformer models and 4 U2++ models, trained with different configurations of data.  Other fusion tricks include LM rescoring for single speaker and multi-speaker ASR models (team G34) and model averaging from different training stages (team B24).


\vspace{-0.4cm}
\section{Conclusions}
\vspace{-0.3cm}

This paper briefly describes the setup of the ICASSP 2022 multi-channel multi-party meeting transcription challenge (M2MeT) and summarizes the outcomes of the challenge, highlighting the major techniques used by the top performing teams. We conclude this paper with listing the following major findings. With limited a mount of data to train systems, data augmentation and simulation are effective for both speaker darization and multi-speaker ASR. Likewise, system fusion is another important trick with steady performance gain if system computational resource is not constrained. Front-end processing techniques are also beneficial for far-field scenarios including meeting transcription -- the task at hand. But uniquely for meetings like the AliMeeting data, speaker overlap should be explicitly addressed. For speaker diarization, TS-VAD is still the superior approach to handle speaker overlap. By using the above-mentioned methods and tricks, the diarization error rate has been lowered to 3\% on AliMeeting. For multi-speaker ASR, Conformer is still the state-of-the-art (single-speaker) ASR model used by most teams and Serialized Output Training is the easy-to-use approach to explicitly consider speaker overlap. 
Front-end and back-end joint modeling using neural networks is also a promising solution that deserves future investigation. The best performing system in the ASR track achieves 18.79\% character error rate given the limited training data.

\vspace{-0.4cm}
\begin{spacing}{0.01}
\scriptsize
\bibliographystyle{IEEEbib}
\bibliography{strings,refs_s}
\end{spacing}

\end{document}